\documentclass[11pt,a4paper]{article}
\usepackage{cite}
\usepackage{amsmath, amsthm, amssymb,slashed,multicol,booktabs, multirow}
\usepackage{yfonts}

\usepackage{ifpdf}
\ifpdf
  \usepackage[pdftex]{graphicx}
  \usepackage{epstopdf}
\else
  \usepackage[dvips]{graphicx}
\fi

\usepackage{subfigure}
\usepackage{xfrac}  
\usepackage{fancyhdr}
\usepackage{mathbbol}
\usepackage{url}
\usepackage{color}
\usepackage{bbm}
\usepackage{rsfso}
\usepackage{dsfont}
\usepackage{bm}
\usepackage{amsfonts}
\usepackage{mathrsfs}
\usepackage{slashed}
\usepackage{wasysym}
\usepackage{yfonts}
\usepackage{ytableau}
\usepackage{tikz}
\usetikzlibrary{arrows,decorations.pathmorphing,hobby}
\usetikzlibrary{calc,arrows.meta}
\usepackage{cancel}
\usepackage[normalem]{ulem}
\def\bl{\begin{equation}\begin{aligned}}
\def\el{\end{aligned}\end{equation}}
\def\beal{\begin{align}}
\def\eal{\end{align}}
\def\be{\begin{equation}}
\def\ee{\end{equation}}
\def\bpm{\begin{pmatrix}}
\def\epm{\end{pmatrix}}
\def\bsm{\begin{bmatrix}}
\def\esm{\end{bmatrix}}
\def\bvm{\begin{vmatrix}}
\def\evm{\end{vmatrix}}
\def\bVM{\begin{Vmatrix}}
\def\eVM{\end{Vmatrix}}
\def\bea{\begin{eqnarray}}
\def\eea{\end{eqnarray}}

\def\1{{\bf 1}}
\def\2{{\bf 2}}
\def\3{{\bf 3}}
\def\4{{\bf 4}}

\def\hat{\widehat}

\usepackage{indentfirst}

\newcommand{\OMIT}[1]{}

\usepackage{mathtools}

\begin{document}
\title{About All-Charm Tetraquarks in S and P Waves}

\author{Hugo Garcia Tecocoatzi$^{*}$, Luciano Maiani$^{**}$, Elena Santopinto$^{+}$, Veronica Riquer$^{**}$,\\  Ailer Rivero-Acosta$^{+,}$
 \\
 {\it $^{*}$ Tecnologico de Monterrey, Mexico.}
  \\
 {\it $^{**}$ Universit\`a di Roma la Sapienza, INFN Sezione Roma1, Italy,}
  \\   {\it $^{+}$ INFN,  Sezione di Genova, Italy.}
  }
  
\maketitle

\begin{abstract}

 We present a quantitative analysis of masses and decays of $cc \bar c \bar c$  tetraquarks, with diquarks and antidiquarks in S and P wave. We study explicitely the decays of the $4P-J^{PC}=3^{--},~2^{--}$ into 4 muons and into hidden- and open-charm mesons and estimate its width and branching ratios.
Based on our results and with the present sensitivity, LHCb, CMS and ATLAS should detect the  $4P-J^{PC}=3^{--},~2^{--}$ fully-charmed tetraquark. We note that one  decay channel of the $4P-J^{PC}=2^{--}$could lead to the identification of the (still controversial) $\chi_{c1}(2P)$. 
    
\end{abstract}

\section{Introduction}  \label{introd}

The observation of exotic hadrons promises to be a crucial step in our understanding of quantum chromodynamics in the non-perturbative regime. Since the discovery of X(3872)~\cite{Belle:2003nnu}, numerous tetraquark candidates have been reported~\cite{ParticleDataGroup:2026aaa}, yet their internal structure remains under discussion, see  e.g,~\cite{Santopinto2025,Esposito:2016noz}, with two main interpretations: compact four-quark states~\cite{jaffe,Maiani:2004vq} and references therein, or two meson molecules, see ~\cite{Guo:2017jvc}  and references therein.
\\

 Between 2020 and 2026, different experiments at the LHC \cite{LHCb:2020bwg, ATLAS:2023bft, CMS:2023owd,CMS:2026tiu}, LHCb,  ATLAS and CMS, reported the observation of three tetraquarks made by two charm and two anticharm quarks, denoted as all-charm tetraquarks, X(6900), X(6600) and X(7100). The first was discovered by LHCb \cite{LHCb:2020bwg} and later confirmed by ATLAS \cite{ATLAS:2023bft} and CMS \cite{CMS:2023owd}, the latter identified by CMS \cite{CMS:2023owd, CMS:2026tiu}, see Tab.~1. 
 \\

 In 2025 CMS determined spin and parity of the observed all-charm tetraquarks finding $J^{PC}=2^{++}$ for all\cite{CMS:2025fpt}. Earlier theoretical articles\cite{Becchi:2020mjz, Becchi:2020uvq} had anticipated that $2^{++}$ S-wave states would be the most visible states, as confirmed by a NRQCD production calculation at order ${{\alpha}_s}^{5}$ in \cite{Belov:2024qyi}.
 \\
 
In this article, we present a study of mass spectrum and decay modes of all-charm P-wave tetraquarks. In particular, we give decay rates and branching ratios of  $4P-J^{PC}=3^{--}, 2^{--}$ states. 

We will discuss an interesting decay of the $4P-$ $J^{PC}=2^{--}$ that could involve the true $\chi_{c1}(2P)$, a "possible competitor" of the X(3872).

\begin{table*}[htb]
\caption{ Measured masses and widths of the three fully charmed tetraquarks from the fits to the $J/\psi J/\psi$ and $J/\psi \ \psi(2S)$ mass spectra from the Run~2+3 data set \cite{CMS:2026tiu}.
    The two uncertainties are statistical (first) and systematic (second).
    For comparison, the results from Run~2~\cite{CMS:2023owd} are also shown.
    Masses and widths in MeV.}
\label{Tab:comparison_2_3_simul}
\centering
    \renewcommand\arraystretch{1.3} 
    \begin{tabular}{ c c c c c }
     & & X(6600) & X(6900) & X(7100) \\
    \hline            
    \multirow{2}{*}{$J/\psi J/\psi$: Run~2~\cite{CMS:2023owd}} 
     & Mass  & $6638_{\,-\,38\,-\,31}^{\,+\,43\,+\,16}$ & $6847_{\,-\,28\,-\,20}^{\,+\,44\,+\,48}$ & $7134_{\,-\,25\,-\,15}^{\,+\,48\,+\,41}$ \\
     & Width & $440_{\,-\,200\,-\,240}^{\,+\,230\,+\,110}$ & $191_{\,-\,49\,-\,17}^{\,+\,66\,+\,25}$ & $97_{\,-\,29\,-\,26}^{\,+\,40\,+\,29}$ \\[4pt]
    \multirow{2}{*}{$J/\psi J/\psi$: Run~2+3} 
     & Mass  & $6593_{\,-\,14}^{\,+\,15} \pm 25$ & $6847 \pm 10 \pm 15$ & $7173_{\,-\,10}^{\,+\,9} \pm 13$ \\
     & Width & $446_{\,-\,54}^{\,+\,66} \pm 87$ & $135_{\,-\,14}^{\,+\,16} \pm 14$ & $73_{\,-\,15}^{\,+\,18} \pm 10$ \\[4pt]
    \multirow{2}{*}{$J/\psi \ \psi(2S)$: Run~2+3} 
     & Mass  &  & $6876_{\,-\,29\,-\,112}^{\,+\,46\,+\,111}$ & $7169_{\,-\,52\,-\,72}^{\,+\,26\,+\,80}$ \\
     & Width &  & $253_{\,-\,100\,-\,134}^{\,+\,290\,+\,140}$ & $154_{\,-\,82\phantom{1}\,-\,181}^{\,+\,110\,+\,140}$ \\
    \end{tabular}

\end{table*}
\section{Model}
Following the CMS study of the observed  lowest-lying $S-$wave $J^{P C}=2^{++ }$ all-charm  tetraquarks \cite{CMS:2025fpt, CMS:2026tiu}, we discuss here mass spectrum and decay modes of P-wave states. 

 On the same line as  Refs.\cite{Becchi:2020mjz, Becchi:2020uvq}, we start with the simplest picture of point-like diquark-antidiquark with color $\bf{3}$ and ${\bf{\bar 3}}$, spin $s=1$\footnote{Spin 1 for both diquarks and antidiquarks is required by Fermi statistics.}, mass $M_{cc}$, bound into color singlet states by the Cornell  potential: 
\be
V=-\frac{4}{3}\frac{\alpha_s}{r} + kr + 2M_{cc}    \label{cornell} \ ,
\ee

We take $\alpha_s=0.33$ and $k=0.18~ \text{GeV}^2$, from Ref.~\cite{Maiani:2024ocb} and include the lowest order relativistic corrections to the potential in Eq.~\eqref{cornell}, as detailed in Appendix \ref{cornellpot}, see also~\cite{Voloshin:2007dx,Soni:2017wvy,Maiani:2024ocb} and references therein.


We leave $M_{cc}$ as a free parameter, starting from the value $M_{cc}=2M_c=2\times 1.317$ GeV, with $M_c$ the charm quark mass that, in a similar scheme, fits the mass of charmonia as reported in Refs.~\cite{Soni:2017wvy, Maiani:2024ocb}. The value  $M_{cc}= 2.894$ GeV ($M_{c}=1.447$ GeV) reproduces the mass of the observed  $2^{++}$ $X(6900)$  \cite{LHCb:2020bwg, ATLAS:2023bft, CMS:2023owd, CMS:2025fpt, CMS:2026tiu} as $L=0$ state with the two diquark spins coupled to total spin $S=2$ and principal quantum number $n=3$. At the same time, it provides a reasonable fit to the masses of the $X(6600)$ and $X(7100)$ GeV with  $n=2, L=0$  and  $n=4, L=0$ states, respectively, Tab.~2.

\begin{table}[ht]
\centering
    \begin{tabular}{|c|c|c|c|c|c|}
   \hline 
{\footnotesize{ $M_ {c}$ }} & {\footnotesize{ $M_{cc}$} } & {\footnotesize {M(1S)}} & {\footnotesize{M(2S)}} & {\footnotesize{M(3S)}} & {\footnotesize{M(4S)}}\\
 \hline
\hline
  {\footnotesize {$1.317$} }& {\footnotesize {$2.634$} } & {\footnotesize {$5.560$} } & {\footnotesize $6.055$} & {\footnotesize {$6.422$} } & {\footnotesize {$6.735$} } \\
  \hline
{ {\footnotesize {$1.417$} }}& {\footnotesize {$2.834$} } & {\footnotesize {$5.935$} } &{ {\footnotesize $6.424$} }&{{\footnotesize {$6.785$} }} &{ {\footnotesize {$7.092$} }}\\
  \hline
 {\footnotesize {${\bf1.447}$}} & {\footnotesize {${\bf 2.894}$} } & {\footnotesize {$6.047$} } & {\footnotesize ${\bf 6.536}$} &{\footnotesize {${\bf 6.895}$} } & {\footnotesize {${\bf 7.200}$} }\\
  \hline
 {\footnotesize {$1.517$} }& {\footnotesize {$3.034$} } & {\footnotesize {$6.311$} } & {\footnotesize $6.796$} & {\footnotesize {$7.151$} } & {\footnotesize {$7.453$} }\\
   \hline
\hline 
\end{tabular}
 \caption {\footnotesize {
 Theoretical masses, $J^{PC}=2^{++}, n=1, 2, 3, 4, L=0$, for different values of $M_{cc}$.}}
 \label{dicharmonia2}
 \end{table}
 
  \begin{figure}[htb!]
   \centering
 \begin{center}
     \includegraphics[width=12.0truecm,angle=-90]{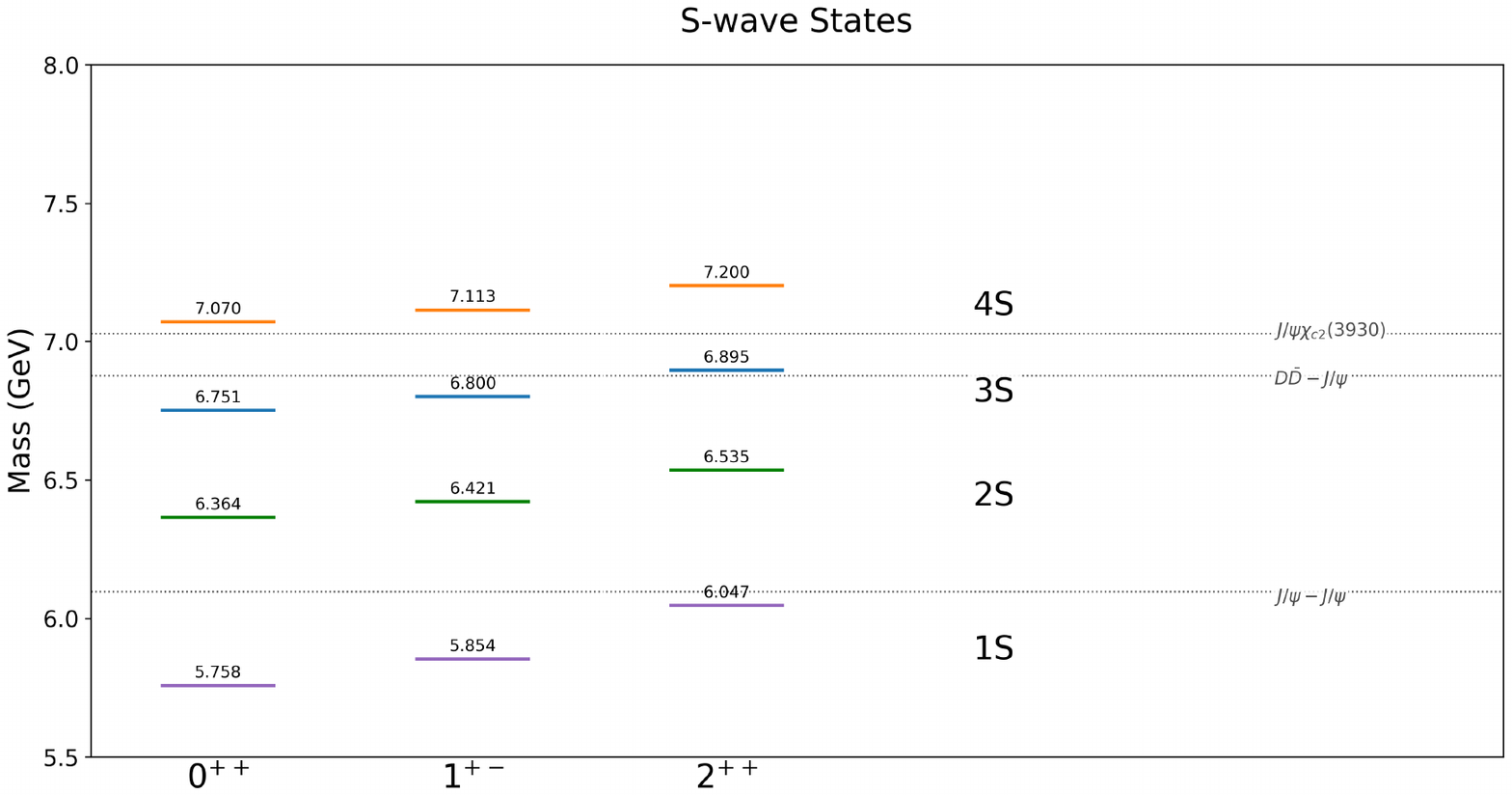}
     \caption{\footnotesize{Theoretical} masses of S-wave fully charmed tetraquarks.}
 \label{UnoS}
  \end{center}
\end{figure}
 

  \begin{figure}[htb!]
   \centering
 \begin{center}
     \includegraphics[width=12.0truecm,angle=-90]{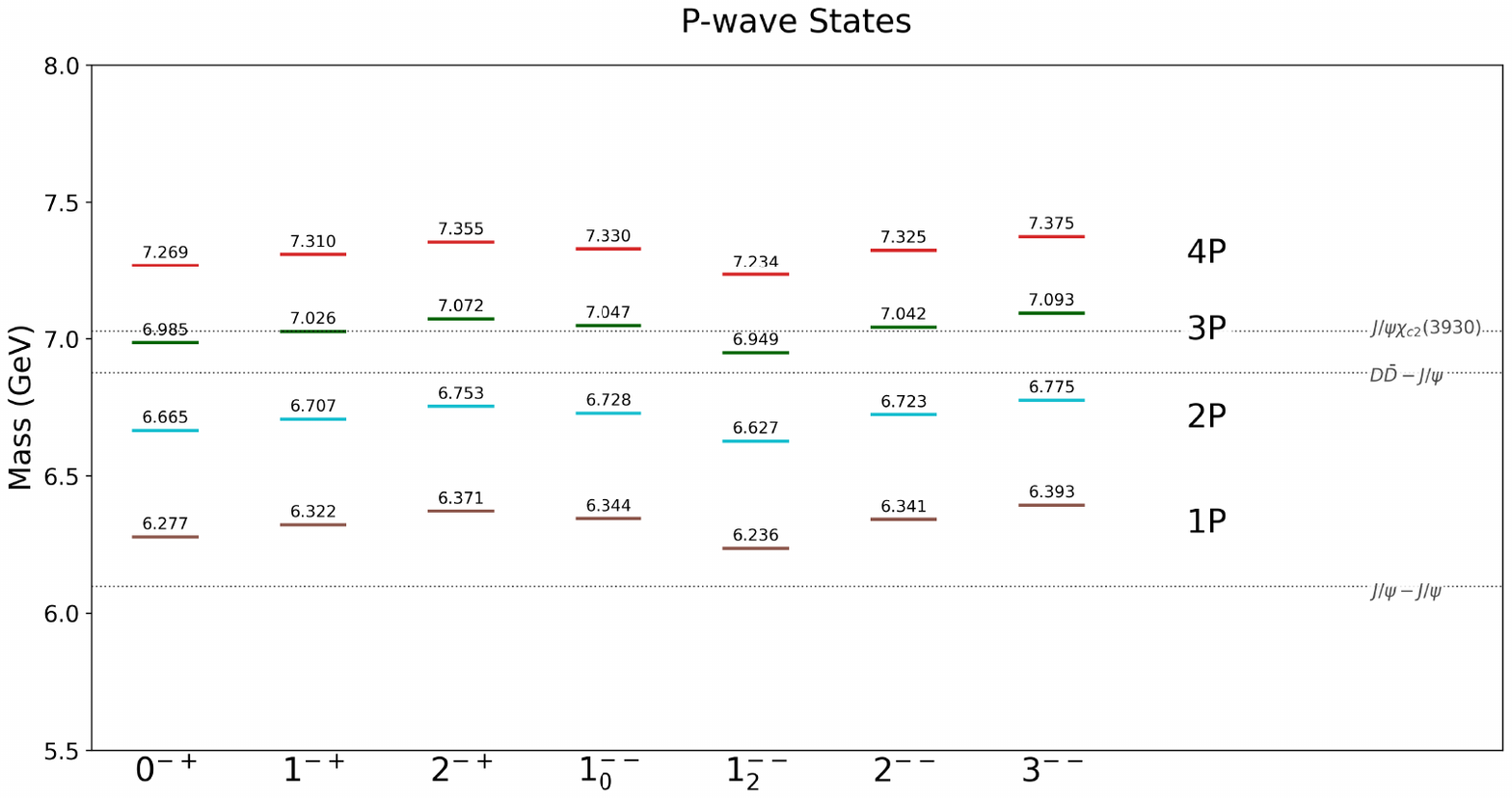}
     \caption{\footnotesize{Theoretical masses of P-wave fully charmed tetraquarks vs. $J^{PC}$ and principal quantum number $n$. Subscripts 2 and 0 indicate the value of the total diquark-antidiquark spin, S. 
     }}
 \label{UnoP}
  \end{center}
\end{figure}
 

Since diquarks and  antidiquarks have relative parity +1, parity and charge conjugation of Dicharmonia are given by:
  \be
  P=(-1)^{L},~C=(-1)^{(L+S)} \ .
  \ee
S being the total spin of the tetraquark.

For $S$-wave states: $J^{PC}=2^{++},1^{+-},0^{++}$. 

For $P$-waves:
\bea
&& (L=1, S=2),~ J^{PC}= 3^{--},~2^{--},~1^{--}  \nonumber \\
&& (L=1, S=1),~ J^{PC}= 2^{-+},~1^{-+},~0^{-+}  \nonumber\\
&& (L=1, S=0),~ J^{PC}=  1^{--}         \label{pstates}    \label{pwave}
\eea

From Eq.\eqref{cornell} we  compute tetraquark masses via the spin-independent Schr\"odinger equation solved numerically, see Ref. \cite{schroed}. Results are reported in Figs.~\ref{UnoS} and \ref{UnoP}. 
Radial wave functions are characterised by the number of nodes, $n_r$, and the orbital angular momentum, $L$. Following the terminology of charmonium spectroscopy, we classify  masses and solutions by the principal quantum number  $n=n_r+1$
~ and  the orbital quantum number $L$, according to the notation:
\be
~\text{Spectroscopic notation of  all-charm tetraquarks}: n^{(2S+1)}L_J \label{spectro}
\ee
 $S$ is the total tetraquark spin and $J$ is the total angular momentum, $J=S+L$.

  \begin{figure}[htb!]
   \centering
 \begin{center}
   \includegraphics[width=12.0truecm]{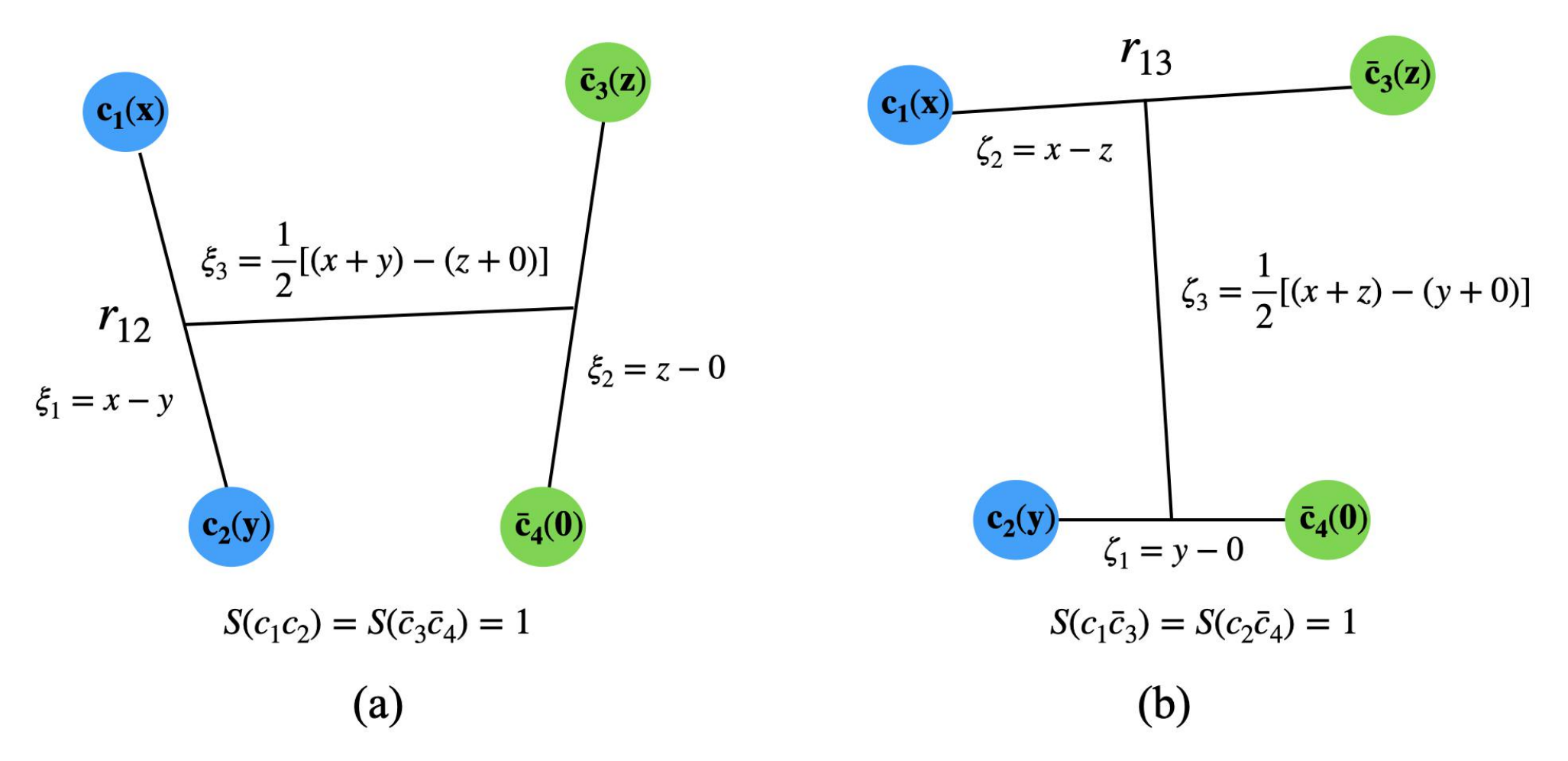}
\caption{\footnotesize {Jacobi coordinates describing the motion of all-charm tetraquark components, Refs.\cite{Anwar:2017toa,Wang:2021kfv}.  Quark and antiquark coordinates are denoted by $\bf{x,~y,~z,~0}$. The two equivalent sets of Jacobi coordinates are denoted by $\xi$, in (a), $\zeta$, in (b). $S(cc),~S(c \bar c)$ denote the spin of quark-quark or quark-antiquark pairs.  Figures from Ref.~\cite{Wang:2021kfv}. } }
\label{Jacobi}
\end{center}
\end{figure}

\section{P-waves with Jacobi coordinates } \label{jacobi}

A crucial parameter in our previous discussion of the decays of all-charm, $S$-wave tetraquarks~\cite{Becchi:2020mjz,Becchi:2020uvq} is the $c\bar c$ overlap probability in the tetraquark, $|\Psi_n(0)_{T}|^2$. The overlap  probability vanishes, however, in the approximation of pointlike diquark and antidiquark in $P$-wave. To describe the decays of $P$-wave tetraquarks, we have to allow for the relative motion of the individual quark and antiquark components, illustrated  in Figs.~\ref{Jacobi}, (a) and (b) .

The full four-body motion of all-charm tetraquarks is discussed in Refs.~\cite{Anwar:2017toa,Wang:2021kfv}, in terms of three Jacobi coordinates $\xi_{1,2,3}$ defined in Fig. \ref{Jacobi}(a), and the spectrum is obtained numerically in Ref.~\cite{Wang:2021kfv} with the complex scaling method \cite{Hiyama:2003cu}.  
In the following, we parametrize  the four-quark wave functions according to (see Appendix~\ref{spatialA}): 

\begin{equation}
    \Psi_{n_1 ,n_{r_1},L_1}( \xi_1,\beta_a)\Psi_{n_2,n_{r_2},L_2}( \xi_2,\beta_b)\Psi_{n_3,n_{r_3},L_3}( \xi_3,\beta)
\end{equation}
where:

\begin{equation}
    \Psi_{ n_i,n_{r_i},L_i}( \xi,\beta)=
\mathcal{N}_{n_{r_i},L_i} e^{-(\beta \xi)^2} \mathcal{L_i}^{L_i+1/2}_{n_{r_i}}(\xi)\mathcal{Y}_{L_i}(\xi)
\end{equation}
  $\mathcal{N_i}_{n_{r_i},L_i}$
 is a normalization factor, $\mathcal{L_i}^{l_i+1/2}_{n_{r_i}}(\xi_i)$  are the appropriate  generalized Laguerre polynomial, $\mathcal{Y}_{L}(\xi)$ the spherical harmonics. The exponents of the Gaussians are obtained from the root-mean-square radii reported in Table VI of Ref.~\cite{Wang:2021kfv}




According to Fig.~\ref{Jacobi} (a), there are three kinds  of $P$-wave states, classified in~\cite{Wang:2021kfv} as:
\begin{itemize}
\item  $\lambda$ states: $L=1$ on $\xi_3$ only;  
\item $\rho$ states: $L=1$ on $\xi_1$ or $\xi_2$, the two sequences being related by charge conjugation. 
\end{itemize}

The results on all-charm  tetraquark masses shown in Fig.~\ref{UnoP}, have  been obtained in the diquark-antidiquark approximation: $\xi_1=\xi_2=0$ and $L=1$ in $\xi_3$ \footnote{A comparison of predicted masses in Figs.~\ref{UnoS} and \ref{UnoP} with those obtained using Jacobi coordinate models is as follows.
Wang et al. \cite{Wang:2021kfv} have more S states  with respect to the diquark antidiquark model since a dicharmonium model can have nodes  only in $\xi_3$ while the 4-body model can have radial excitations also in $ \xi_1 $ and $\xi_2$ coordinates ($\rho$ modes).
$2S$-wave state masses of Ref.\cite{Wang:2021kfv} are in reasonable agreement with the $3S$ masses of Fig. \ref{UnoS}. However, $ 2P~\lambda$ masses of Ref.\cite{Wang:2021kfv} are some $300$ MeV lighter than the $3P$ masses of Fig. \ref{UnoP}.}. As we show below, due to charge conjugation invariance, the $\lambda$-mode on $\xi_3$ is in fact the only possible configuration for $L=1$, $3^{--},~2^{--},~1^{--},~S=2$ states, even when we consider 4 quark configurations, see~\cite{Wang:2021kfv}.

The P-wave 4 quark $\lambda/\xi_3$ wave functions can be  written as:
\bea
&&\Psi_{1,0,0}(\xi_{1},\beta_{a})\Psi_{1,0,0}(\xi_{2},\beta_{b})\Psi_{n,n_r,1}( \xi_3,\beta)= \nonumber\\
&&=\frac{\mathcal{N}_{n_r,1}}{\pi}  \left(\frac{\beta^2_a\beta^2_b}{\pi^{2/3} }\right)^{3/4} e^{-\beta^2_a \xi^2_{1}/2-\beta^2_b \xi^2_{2}/2} e^{-(\beta \xi_3)^2/2} \mathcal{L}^{3/2}_{n_r}(\xi_3)\mathcal{Y}_{1}(\xi_3)         \label{parametriz}
\eea

To bring $c$ and $\bar c$ in the same bilinear and compute the quantity of interest for the decays, i.e. the $\bar c c$ overlap probability $|\Psi_{n}(0)_T|^2$ (see Appendix C for its definition) we have to perform a Fierz transformation, which amounts to pass from diagram (a) to diagram (b) in Fig.~\ref{Jacobi}. After the coordinate transformation, we still have  diquarks and antidiquarks   in spin-1, combined to $S=2$ and  we have to perform a further Fierz transformation on spin variables. One can show that the result has both $(\bar c c)$ combinations in spin 1 and total spin $S=2$.

Coordinate transformation of $\lambda$ P-wave states give rise, in the meson-meson basis, to the superposition of two $L=1,~\rho^\prime$ states: $(\bar c_3 c_1)^1(\bar c_4 c_2)^{1\times P-wave}$ and $(\bar c_3 c_1)^{1\times P-wave}(\bar c_4 c_2)^1$  diagram \ref{Jacobi} (b), which however differ only for the exchange of identical particle (see Appendix~\ref{spatialA}). 
In addition, the $\lambda^\prime$ configuration in (b) is forbidden, since the $L=1$ harmonic antisymmetrizes $(\bar c_3 c_1)^1$ and $(\bar c_4 c_2)^1$, which commute.

\section{Decays of P-wave tetraquarks}

We consider the three $P-$ wave states $4^{S=2} P_{J=3,2,1}$ reported in Fig.~\ref{UnoP}, see Eq.~\eqref{pstates}, which we denote as:  
\be
\chi_{cc3}(7.375),~\chi_{cc2}(7.324),~\chi_{cc1}(7.234)
\ee

They originate from the state: 
\be
\left[\Big | [(cc)^1_{\bar 3}, (\bar c\bar c)^{1}_3]^{2,m_s} \Big>_1
\otimes  \Big|[n=4, L=1,m_l\Big>_{\rm spatial}\right]^J_{M} =\Big | [(cc)^1_{\bar 3}, (\bar c\bar c)^{1}_3]^{2\times {\text{P-wave}}} ]_1^3\Big>
\ee

with  $P$-wave  diquark-antidiquark, total spin $S=2$ and $J=3 ,2, 1$ respectively. In the following we shall understand always M=J. 

We consider first $3^{--},~n=4$. Using spin-color Fierz and spatial transformations we find, in the meson-meson basis of Figure \ref{Jacobi}(b):
\be 
\Big | \big [[(cc)^1_{\bar 3}, (\bar c\bar c)^{1}_3]^{2 \times \text{P wave}} \big]_1^3\Big>=  -\sqrt{\frac{1}{3}}\Big |\big [[(\bar c c)^1_1, (\bar c c)_1^{1\times \text{P wave}]}\big]^3_1 \Big >+  \sqrt{\frac{2}{3}} \Big |\big [[(\bar c c)^1_8, (\bar c c)_8^{1\times \text{P wave}]}\big]^3_1 \Big >
 \label{colorP}
\ee 
Similar equations hold for $J^{PC}=2^{--},~1^{--}$.

  \begin{figure}[htb!]
   \centering
 \begin{center}
   \includegraphics[width=10.0truecm]{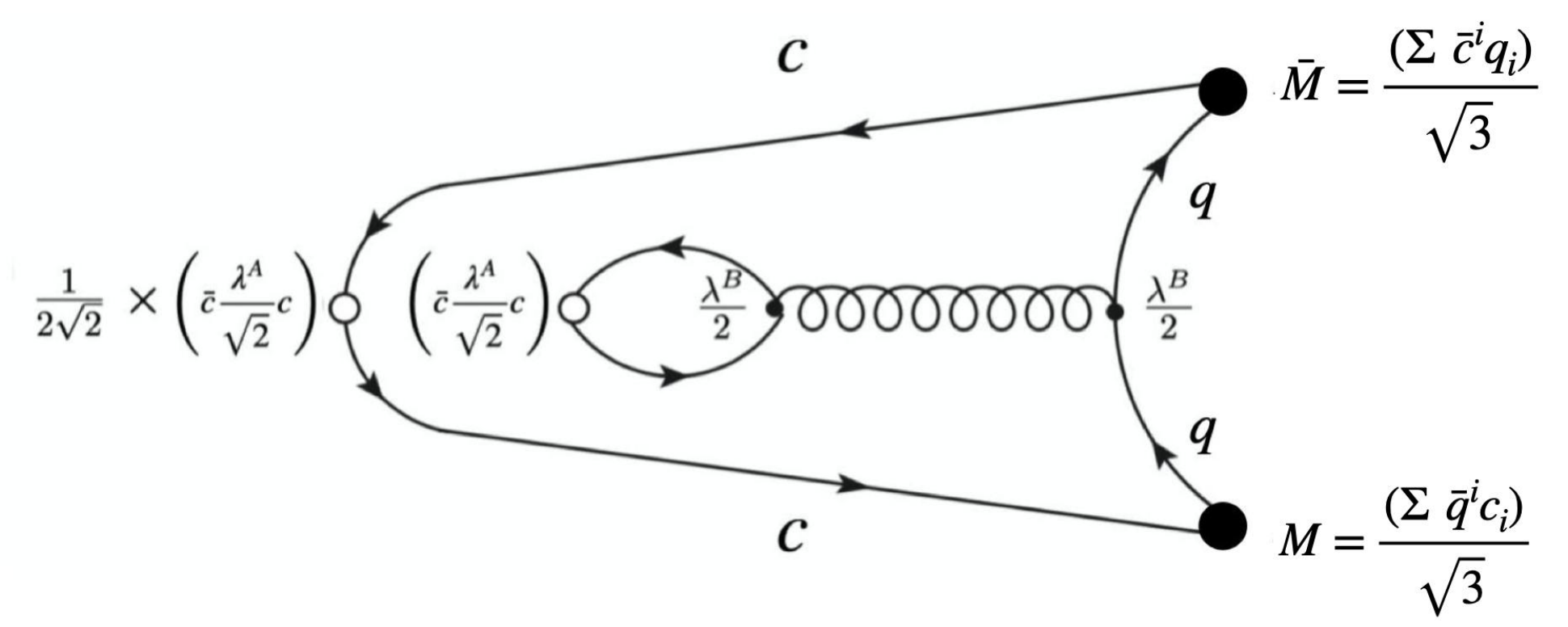}
 \end{center}
\caption{\footnotesize {Decay of the octet$\times$ octet component of the P-wave tetraquark, diagram from Refs. \cite{Becchi:2020mjz,Becchi:2020uvq}. The external $c\bar q$ and $\bar c q$ mesons are represented by normalised color singlets. The Chan-Paton factor in  Eq.~\eqref{rate3} is the product of the traces of the two loops. }}
\label{Tre}
\end{figure}

The first term of the r.h.s. of~\eqref{colorP} gives rise to the decay: 
\be
\chi_{cc3}(7.375)\to J/\Psi+\chi_{c2}(3.930) \label{decay1}
\ee
   
   The  contribution of the octet $\times$ octet component of \eqref{colorP} is described by the diagram of Fig.~\ref{Tre} \cite{Becchi:2020uvq}, which produces four possible final states $(\bar c q)(\bar q c)$ ($q=u, d, s, c$).
   
   \begin{itemize}
   \item For $q=c$ we obtain a further, dominant, contribution to the decay Eq.~\eqref{decay1};
   \item For $q=u, d, s$ we obtain decays into $D^*$(S-wave)$\bar D^*$(P-wave) or $D_s^*$(S-wave)$\bar D_s^*$(P-wave).
   \end{itemize}

 \section{Rates}

 We define:
 \be
  \xi_n=\frac{|\Psi_n(0)_{ T}|^2}{|\Psi(0)_{J/\Psi}|^2}  \label{xival}
  \ee 
where $|\Psi_n(0)_{T}|^2$ is the overlap probability of the $c\bar c$ pair in the P-wave tetraquark with principal quantum number n. The calculation of  $|\Psi_{n}(0)_{T}|^2$ is presented in Appendix \ref{numerical}. Numerical values of $|\Psi_n(0)_{T}|^2$ (GeV$^3$) for $\lambda~P$-wave state with $J=3^{--}$ are given in Tab.~\ref{overlap}, derived from the values of radii reported in Table 6 of Ref.\cite{Wang:2021kfv}. Using the orthonormality of the spherical armonics, one shows easily that the same results for $|\Psi_n(0)_T|^2$  apply to  $J=2^{--}, 1^{--}$.

The $c\bar c$ overlap probability n i$J/\Psi$, $|\Psi(0)_{J/\Psi}|^2$, is determined by the formula: 
\be
\Gamma(J/\Psi \to \mu^+\mu^-)=Q_c^2 ~\frac{16 \pi \alpha^2}{3} ~\frac{ |\Psi_ {J/\Psi}(0)|^2}{m^2_{J/\Psi}}
\label{VRW}
\ee
proposed in Ref.~\cite{VanRoyen:1967nq} for $\rho_0 \to \mu^+ \mu^-$. Using the experimental  $J/\Psi$ decay rate one obtains:
\be
 |\Psi_ {J/\Psi}(0)|^2=0.13~\text{GeV}^3
 \ee
and, with Table~\ref{overlap}, we find:
 \be
 \xi_4 \simeq 0.04 \label{value}
 \ee

 \begin{table}[htb!]
\centering
    \begin{tabular}{|c|c|c|c|c|}
     \hline
{\footnotesize Principal Quantum Number} &{\footnotesize $n=1$}&{\footnotesize $n=2$}&{\footnotesize $n=3$}&{\footnotesize $ n=4$}\\ \hline 
{\footnotesize{ $|\Psi_n(0)_{T}|^2$~(GeV$^3$)} }& {\footnotesize $0.0089$} &{\footnotesize $0.0066$} & {\footnotesize $0.0054$}& {\footnotesize $0.0048$}\\
 \hline
\end{tabular}
  \caption{ \footnotesize {$\bar c c$ overlap probabilities (GeV$^3$) for $\lambda~P$-wave states vs.~$n$.}}  
  \label{overlap}
\end{table}

   	Rates are computed following Refs.~\cite{Becchi:2020mjz,Becchi:2020uvq}  \footnote{ In the following equations: 
  $\frac{1}{3}$ or  $\frac{2}{3}$ from the spectroscopic factors in Eq.~\eqref{colorP}. 
 The amplitude corresponding to the diagram of Fig.~\ref{Tre}~\cite{Becchi:2020mjz,Becchi:2020uvq} is of order $\alpha_s$ times the trace of the two loops, sometimes called the Chan-Paton factor:
   \be
   A=\alpha_s\frac{1}{2\sqrt{2}}\text{Tr} \big[\frac{\lambda^A} {\sqrt{2}} ( \frac{1}{\sqrt{3}}) \frac{\lambda^B}{2} (\frac{1}{\sqrt{3}}) \big]\times \text{Tr} \big[\frac{\lambda^A}{\sqrt{2}}\frac{\lambda^B}{2}\big]=\alpha_s \frac{\sqrt{2}}{3} \label{chanpat}
   \ee
  Sum over repeated indices understood; $Tr(\lambda^A \lambda^B)=2\delta_{AB}; A, B=1, 2,...8$. }
  \begin{itemize}
 \item From the first term in Eq.~(\ref{colorP}):
 \bea
 &&\Gamma_1(cc3)=  \Gamma(\chi_{cc3}(7.380)\to \chi_{c2}(3.930)
 +\text{hadrons~(mostly}~J/\Psi)=\notag \\
 &&=  \frac{1}{3}|\Psi_4(0)_T|^2 v \sigma(c \bar c \to 3~\text{ gluons})= \frac{1}{3}~ \xi_4 \times \Gamma(J/\Psi)=\xi _4\cdot 31~\text{keV} \label{rate1}
 \eea
 
  \item  From the diagram in Fig.~\ref{Tre}:
 \bea 
 && \Gamma_2(cc3)= \frac{2}{3}~\frac{2}{9}~|\Psi_{J/\Psi}(0)|^2~(\frac{16\pi\alpha^2_S}{3} ~\frac{1}{m^2_{J/\Psi}}~)  \times \xi_4 \times 4= \notag \\
 && = \xi_4 \cdot 15~\text{MeV}     \label{rate3}
 \eea
 
  \end{itemize}
The final states in $\Gamma_2( cc3)$ are of the form $M(\bar c q)_{S-wave}\times M(\bar q c)_{P-wave}$ with $q=u,d; s; c$. For $q=c$ we obtain again the final state reported in Eq.~\eqref{rate1}.

The total decay rate is: 
\be
\Gamma(\chi_{cc3}(4P))\simeq  \Gamma_2( cc3)=\xi _4\cdot 15~\text{MeV}\label{totalG}
\ee

Final states and corresponding branching fractions are reported in Tab.~\ref{uno} for different $q=c;u,d; s$.

  \begin{table}[htb!]
\centering
    \begin{tabular}{|c|c|c|c|c|c|}
     \hline
-- &{\footnotesize $J^{PC}$}&{\footnotesize $J/\Psi+\chi_{c2}(2P)(q=c)$}&{\footnotesize $D^*(2007) \bar D_{2}^*(2460)$($q=u,d$)}&{\footnotesize $ D_s^*(2107) \bar D_{s2}^*(2573)$($q=s$)}&{\footnotesize {$\chi_{c2}(2P)+\mu^+\mu^-$}}\\ \hline 
$\chi_{cc3}(4P)$&{\footnotesize $3^{--}$ }&{ \footnotesize $0.25$}&{ \footnotesize $0.50$}&{ \footnotesize $\sim 0.25$}&{ \footnotesize $1.5 ~10^{-2}$}\\
 \hline
\end{tabular}
  \caption{ \footnotesize {Branching fractions of the $J^{PC}=3^{--}$ fully-charmed $P-$ wave tetraquark.}}  \label{uno}
\end{table}  
{ With the appropriate changes, i.e. $\chi_{c2}(2P) \to \chi_{c1}(2P),~ \bar D_{s2}^* \to  \bar D_{s1}^*$ etc., we obtain the same branching fractions for $J^{PC}=2^{--}$ and $1^{--}$. Cross sections for production in $p p$ collisions scale as $2J+1$.

 \section{$\chi_{cc2}(7.324)$: a special case}
 
 Following Eq.~\eqref{colorP} we write, for the $2^{--},S=2,~4P$ state:
\be 
\Big | \big [[(cc)^1_{\bar 3}, (\bar c\bar c)^{1}_3]^{2 \times \text{P wave}} \big]_1^2\Big>=  -\sqrt{\frac{1}{3}}\Big |\big [[(\bar c c)^1_1, (\bar c c)_1^{1\times \text{P wave}]}\big]^2_1 \Big >+  \sqrt{\frac{2}{3}} \Big |\big [[(\bar c c)^1_8, (\bar c c)_8^{1\times \text{P wave}]}\big]^2_1 \Big >
 \label{colorP2}
\ee 
The first term in the  r.h.s. gives rise to the decay:
\be
\chi_{cc2}(7.324)\to J/\Psi + \chi_{c1}(2P) \label{cc2decay}
\ee

The nature of the state $\chi_{c1}(2P) $ is still controversial.
In the PdG list, the charmonium state reported as $\chi_{c1}$ is the former $X(3872)$, the first discovered "exotic hadron".  However, judging from its  $J/\Psi+\rho^0$ decay~\cite{LHCb:2022jez}, $X(3872)$ could well be a tetraquark~\cite{Maiani:2004vq} or a molecule~\cite{Guo:2017jvc,Zhang:2024fxy}. In addition, the prediction for the $2P$, spin 1 charmonium mass, computed with the Cornell potential, is rather $M=3.940-3.950$ GeV~\cite{Soni:2017wvy, Maiani:2024ocb}.

A charmonium explanation of $X(3872)$ has been proposed in \cite{DeFazio:2008xq} on the basis of its radiative decays. For a charmonium state close to a threshold, the downward mass shift could be explained in the unquenched quark model \cite{Ferretti:2013faa, Ferretti:2014xqa} as due to self-energy corrections, an explanation that could also justify the isospin-violating decay $J/\Psi+\rho^0$ as due to its meson-meson virtual components  \cite{Ferretti:2018tco}.

Finally, a particle that could be identified with $\chi_{c1}(2P)$ in Eq.~(\ref{cc2decay}) is the $D^*\bar D$ peak with mass $3940$ MeV, observed  by Belle in $e^+e^-\to J/\psi+\text{hadrons}$~\cite{Belle:2005lik} and classified in PDG as X(3940), without specification of its quantum numbers.

In conclusion, the observation of the decay of $\chi_{cc2}(7.345)$
may lead to the identification of the true $\chi_{c1}(2P)$, an important step in our understanding of hadron spectroscopy.

\section{Acknowledgements}
We thank Liupan An and Vanya Belyaev for illuminating discussions about LHCb results.

\newpage

\appendix
\section{Cornell potential and its relativistic corrections for Dicharmonium} \label{cornellpot}


Following the treatment of the non-relativistic hydrogen atom, we separate angular and radial wave functions for each value of the orbital angular momentum $L$, according to
\be
\psi(n, L,m)(x)=R(n,L)(r) \cdot Y(\theta, \phi)^L_m \ ,
\ee
with $n=1,2,\dots$ the principal quantum number, $R$ the radial wave function and $Y^L_m$ the spherical harmonics (in the standard notation: $L=S,P,D,\dots =0,1,2,$). Further, we put:
\be
R(r)=\frac{\chi(r)}{r} \ , 
\ee
with $\chi(r)$, the reduced radial wave function, satisfying the boundary condition
\be
\lim_{r\to 0^+} \chi(r)=0 \ .
\ee

Similarly to atomic physics, hyperfine, spin-orbit and tensor interactions arise as part of the expansion in powers of $v^2/c^2$ of the relativistic quark-antiquark potential and are to be treated as first-order perturbations to the diquark-antidiquark interaction \eqref{cornell}, see also~\cite{Soni:2017wvy,Voloshin:2007dx}\footnote{A couple of misprints in the formulae of Ref.~\cite{Maiani:2024ocb} are corrected here. } 

\emph{\bf{Hyperfine Interaction.}}
The spin-spin interaction, taken from atomic physics, is:
\bea
&&V_{SStot}=\frac{2}{3 M_c^2}\nabla (V_V) ({{\bf{s \cdot \bar{s}}}})=\frac{16\pi \alpha_s}{9M_{cc}^2} \delta^{(3)}(r)~[S(S+1)-4] \nonumber \\
&&<V_{SStot}>=\frac{16\pi \alpha_s}{9M_{cc}^2}~|\psi(0)|^2~[S(S+1)-4]   \label{hfinter}
\eea
where $S$ is the Dicharmonium total spin and:
\be
|\psi(0)|^2=\frac{1}{4\pi}\lim_{r\to 0^+} (\frac{\chi(r)}{r})^2\neq 0 ~\text{for}~L=0 .
\ee
The hyperfine interaction is non vanishing for $L=0$ states only, and: $J^{PC}=S^{PC}=2^{++},~1^{+-},~0^{++}$.

Defining
\be
{\bar V}_{SS}=\frac{16\pi \alpha_s}{9M_{cc}^2} |\psi(0)|^2  \label{valSS}  
\ee
one has:

\be
<V_{SStot}>_{S-wave}={\bar V}_{SS}\times \Big\{\begin{array}{c} +2,~(S=2)\\-2,~(S=1)\\-4,~(S=0)\end{array} \ .
\ee

\emph{\bf{Spin Orbit Interaction.}}

We define
\bea
&&V_{LS}=\frac{1}{2M_{cc}^2 r}\big(3\frac{d V_V}{dr}-\frac{d V_S}{dr}\big)= \frac{1}{2M_{cc}^2}\big(\frac{4\alpha_s}{r^3}-\frac{k}{r}\big) \ ,
\notag \\
&&V_{LStot}=V_{LS} ({\bf{L\cdot S}}) \ ,
\eea
${\bf S}$ being the total Dicharmonium spin.
To first order in this potential, we compute the integral
\be
{\bar V}_{LS}=\frac{1}{2M_{cc}^2} \int_0^{+\infty}\big(\frac{4\alpha_s}{r^3}-\frac{k}{r}\big)r^2  (\frac{\chi(r)}{r})^2 dr  \ . \label{valLS}
\ee

The integral is convergent at the origin for $L\geq 1$ since $(\chi(r)/r )^2\sim r^{2L}$ for $r\to 0$ and:
\be
\delta m_{LS}=<V_{LStot}>_{P-wave}={\bar V}_{LS}({\bf{L\cdot S}}) =\frac{1}{2}~ {\bar V}_{LS} [J(J+1)-S(S+1)-2] \label{soint}
\ee
For $P$-waves, we report, in the square brackets, the value of the square bracket in \eqref{soint} i.e. the value of $2({\bf L}\cdot{\bf S})$):
\bea
&& (L=1, S=2),~ J^{PC}= 3^{--}~[+4],~2^{--}~[-2],~1^{--}~[-6]  \nonumber \\
&& (L=1, S=1),~ J^{PC}= 2^{-+}~[+2],~1^{-+}~[-2],~0^{-+}~[-4]  \nonumber\\
&& (L=1, S=0),~ J^{PC}=  1^{--}~[0]        
\eea

\emph{\bf{Tensor Interaction.}}

One defines the tensor interaction starting from (see  \cite{Voloshin:2007dx,Soni:2017wvy,Wang:2021kfv})

\bea
&&V_T=\frac{1}{6M_{cc}^2}\big[(\frac{d^2 }{dr^2}-\frac{1}{r} \frac{d }{dr})V_V(r)\big]  \ , \notag \\
&&V_{Ttot}=-3V_T [N_{ij} S_i S_j ]=\frac{1}{2M_{cc}^2} \big(\frac{4\alpha_s}{r^3}\big)[N_{ij} S_i S_j ] \ ,  
\label{tensor}
\eea

\be
N_{ij}=(\hat {n}_i\hat {n}_j-\frac{1}{3} \delta_{ij}),~\hat{n}_i=\frac{r_i}{r} \ . \label{hat}
\ee

The average value of $N_{ij}$ in a state with definite orbital angular momentum $L$ is (Landau and Lifshitz~\cite{Landau:1977}
\footnote{The r.h.s of \eqref{tensor2} is the only symmetric and traceless tensor with two indices, which can be formed with $L_i$, the only vector remaining after integration over $r$. To obtain \eqref{const}, one multiplies both sides of \eqref{tensor2} by $L_i$ (on the left) and $L_j$ (on the right), summing over $i$ and $j$.  The result follows from the fact that $\hat{n}$ is orthogonal to $L=\vec{r}\times\vec{p}$ and from angular momentum commutators.}):
\bea
&&<(\hat {n}_i\hat {n}_j-\frac{1}{3} \delta_{ij})>_L= a(L)\big[L_i L_j+L_j L_i-\frac{2}{3} \delta_{ij}L(L+1)\big] \ , \label{tensor2} \\
&&a(L)=-\frac{1}{4L(L+1)-3} \ , \label{const}
\eea
so that:
\bea
&&<V_{Ttot}>_L=<\frac{1}{2M_{cc}^2}\big(\frac{4\alpha_s}{r^3}\big)>_{L} \times \nonumber \\
&&\times 2a(L)\big[ ({\bf S} {\bf \cdot} {\bf L})^2-\frac{1}{3}S(S+1)L(L+1)\big] \ .
\eea
The effect, of course, vanishes for $L=0$ or $S=0$. For $L=1$, we compute the integral
\bea
&&V_1=<\frac{1}{2M_{cc}^2}\big(\frac{4\alpha_s}{r^3}\big)>_{P-wave} = \int_0^\infty  \frac{1}{2M_{cc}^2} \big(\frac{4\alpha_s}{r^3}  \big) r^2  (\frac{\chi(r)}{r})^2 dr \ ,
\eea
and find
\bea
&&\delta m_T= <V_{Ttot}>_{P-wave}={\bar  V}_T \cdot T  \notag \\
&& {\bar  V}_T=\frac{1}{2}\frac{1}{M_{cc}^2}<\big(\frac{4\alpha_s}{r^3}\big)>_{P-wave}\ . \label{valT} \\
&& T=-\frac{2}{15}~[\frac{3}{4} B^2-2 S(S+1)] \label{Tenscoeff}
 \eea
 
 where $B$ indicates the values of the square bracket in \eqref{soint}, reported in \eqref{pstates}. For the seven $P-$wave states listed in \eqref{pstates}, we find:
 \bea
 &&  T=\Big \{\begin{array}{c} 0~~~(3^{--})\\ 6/5~(2^{--}) \\ -2~(1^{--} )\end{array};~\Big \{\begin{array}{c} 2/15~~~~(2^{-+})\\  2/15~~~~(1^{-+}) \\ - 16/15~(0^{-+} )\end{array};~~0~(1^{--}). 
  \eea

 \section{Spatial wave-functions} \label{spatialA}


The tetraquark wave function for a \(P\)-wave state in the diquark--antidiquark basis is
\begin{equation}
|3^{--}\rangle_M=
\sum_{m_s,m_l}
\langle 2,m_s;1,m_l|3,M\rangle\,
\Big|[(cc)^1_{\bar 3},(\bar c\bar c)^1_3]^2_{m_s}\Big\rangle_{\rm 1}
\otimes
| n,n_r,L=1,m_l\rangle_{\rm spatial},
\end{equation}
where \(n=n_r+1\). 
Defining the common factor
\begin{equation}
\mathcal N=
\frac{1}{\pi}
\left(\frac{\beta_a^2\beta_b^2}{\pi^{2/3}}\right)^{3/4}
e^{-\beta_a^2\xi_1^2/2-\beta_b^2\xi_2^2/2}
e^{-\beta^2\xi_3^2/2},
\end{equation}
with the Jacobi coordinates \(\xi_i\) defined in Fig.~\ref{Jacobi}, the
spatial wave functions for the first four \(P\)-wave excitations can be parametrized as  in Eq.~\eqref{parametriz}, with  $\beta_{a,b}$ and $\beta$ fixed to reproduce the rootmean square values reported in  Ref. \cite{Wang:2021kfv} Table IV}:

\begin{align}
|n=1,n_r=0,L=1,m_l\rangle_{\rm spatial}
&=
\mathcal N
\left(\frac{8\beta^5}{3\sqrt{\pi}}\right)^{1/2}
\xi_3\,Y_1^{m_l}(\hat\xi_3),
\\[2mm]
|n=2,n_r=1,L=1,m_l\rangle_{\rm spatial}
&=
\mathcal N
\left(\frac{16\beta^5}{15\sqrt{\pi}}\right)^{1/2}
\xi_3
\left(
-\beta^2\xi_3^2+\frac52
\right)
Y_1^{m_l}(\hat\xi_3),
\\[2mm]
|n=3,n_r=2,L=1,m_l\rangle_{\rm spatial}
&=
\mathcal N
\left(\frac{16\beta^5}{105\sqrt{\pi}}\right)^{1/2}
\xi_3
\left(
\beta^4\xi_3^4
-7\beta^2\xi_3^2
+\frac{35}{4}
\right)
Y_1^{m_l}(\hat\xi_3),
\\[2mm]
|n=4,n_r=3,L=1,m_l\rangle_{\rm spatial}
&=
\mathcal N
\left(\frac{\beta^5}{70\sqrt{\pi}}\right)^{1/2}
\frac{\xi_3}{9} 
\left(
-\beta^6\xi_3^6
+\frac{27}{2}\beta^4\xi_3^4
-\frac{189}{4}\beta^2\xi_3^2
+\frac{315}{8}
\right)
Y_1^{m_l}(\hat\xi_3).
\end{align}

\section{Numerical evaluation of the $|\Psi_n(0)_{T}|^2$ probabilities} \label{numerical}

Here we evaluate the  $|\Psi_n(0)_{T}|^2$ probabilities, starting with the $J=3^{--}$ state. First we express coordinates $\xi_i$, graph in Fig.~\ref{Jacobi}(a), in terms of coordinates $\zeta_i$, graph in Fig.~\ref{Jacobi}(b), from the relations:
\bea
&r_{12}=\xi_1=x-y,~r_{34}=\xi_2=z;~r=\xi_3=\frac{1}{2}(x+y-z); \notag \\
&r_{13}=\zeta_1=x-z,~r_{24}=\zeta_2=y;~r^\prime=\zeta_3=\frac{1}{2}(x+z-y) \notag
\eea
that is:
\bea
&\xi_3=\notag\frac{1}{2}(\zeta_1+\zeta_2)  \notag \\
&\xi_1=\frac{1}{2}[(\zeta_1-\zeta_2)+2\zeta_3]\notag \\
&\xi_2=-\frac{1}{2}[(\zeta_1-\zeta_2)-2\zeta_3] \label{trasf}
\eea

\vskip 0.3 cm

We shall take $\beta_a=\beta_b=\beta $. 
Thus, we can parametrize the first P-wave state as:
\bea
|n=1,n_r=0,L=1,m_l\rangle_{\rm spatial}&=\frac{\beta_0^{3}}{\pi^{3/2}}~
\exp\!\left[
-\frac{\beta_0^2}{4}
(\mathbf \zeta_1-\mathbf \zeta_2)^2
-
\beta_0^2\,\mathbf \zeta_3^2
\right] \nonumber \\  &\times\left(\frac{2^{3}\beta^{5}}{3\sqrt \pi}
\right)^{1/2}~(\frac{1}{2})|\zeta_1+ \zeta_2|  Y_1^{m_l}(\hat{\zeta_1+\zeta_2})
\exp\!\left[
-\frac{\beta^2}{8}
(\mathbf \mathbf \zeta_1+\zeta_2)^2
\right] 
\eea

We square the modulus of the wave function of Eq. \eqref{trasf} and, in the limit $\zeta_1 \to 0$, integrate over $\zeta_2$ and $\zeta_3$. Using the overall rotational symmetry, we align $\zeta_3$ with the z-axis (which gives a factor of $4\pi$) and integrate over the possible orientations of $\zeta_2$,  to obtain  
\bea
|\Psi_1(\zeta_1=0)_T|^2=\frac{4\beta_0^3\beta^5}{\sqrt{2}~ \pi^{3/2}(\beta^2+2\beta_0^2)^{5/2}}
\eea

 The values od $\beta$ and $\beta_0$, the exponents of the Gaussians, are obtained from the root-mean-square radii reported in Table VI of Ref.~\cite{Wang:2021kfv}:
\be
\beta=0.72 ~\text{GeV},~\beta_0=0.37~ \text{GeV}, r=\frac{\beta_0}{\beta}=0.51
\ee
and find
\be
|\Psi_1(0)|^2= 0.0089~\text{GeV}^3
\ee

Similarly, we find

\begin{itemize}
    \item For 2P
    \be
|\Psi_2(\zeta_{1}=0)_T|^2= \frac{2\sqrt{2}~\beta_0^3 \beta^5}{\pi \sqrt{\pi}} ~ \frac{(10 \beta_0^4+\beta^4)}{(2\beta_0^2+\beta^2)^{9/2}}= 0.0066~\text{GeV}^3
\ee

\item For 3P

\be
|\Psi_3(\zeta_{1}=0)_T|^2=\frac{4 \beta_0^3 }{\pi ^{3/2}~\sqrt{2} }~\frac{\beta^5~\big(70 \beta_0^8+28\beta_0^4\beta^4+\beta^8\big)}{\big(2\beta_0^2+\beta^2\big)^{13/2}}=0.0054~\text{GeV}^3
\ee

\item For 4P, we use: $r= \beta_0/\beta$

\be
|\Psi_4(\zeta_{1}=0)_T|^2= \frac{4\beta_0^3}{\pi^{3/2}  \sqrt{2}}~\frac{\big(420 r^{12}+378 r^8 + 54 r^4 +1\big)}{\big( 1 + 2 r^2\big)^{17/2}}=0.0048~\text{GeV}^3
\ee

The same results are obtained for $2^{--}$ and $1^{--}$ states, using the orthonormality of the spherical wave functions.

\end{itemize}

  
\end{document}